\documentstyle[aps,prb,preprint]{revtex}
\input psfig.tex
\begin{document}
\author{F. Calvo \\ D\'epartement de Recherche sur la Mati\`ere Condens\'ee, \\
Service des Ions, Atomes et Agr\'egats, CEA Grenoble \\
F38054 Grenoble Cedex, France \\ $ $ \\
J. P. Neirotti and David L. Freeman \\
Department of Chemistry, University of Rhode Island\\
51 Lower College Road, Kingston, RI 02881-0809\\
and\\
J. D. Doll\\
Department of Chemistry, Brown University\\
Providence, RI 02912}
\title{Phase changes in 38-atom Lennard-Jones clusters. II:
A parallel tempering study of equilibrium and dynamic properties
in the molecular dynamics and microcanonical
ensembles}
\date{\today}
\maketitle

\begin{abstract}
We study the 38-atom Lennard-Jones cluster with parallel tempering Monte Carlo
methods in the microcanonical and molecular dynamics ensembles.
A new Monte Carlo algorithm is
presented that samples rigorously the molecular dynamics ensemble for a
system at constant total energy, linear and angular momenta. By combining the
parallel tempering technique
with molecular dynamics methods, we develop a hybrid method to overcome
quasi-ergodicity and to extract both equilibrium and dynamical properties 
from Monte Carlo and molecular dynamics
simulations. Several thermodynamic, structural and dynamical properties are
investigated for LJ$_{38}$, including the caloric curve,
the diffusion constant and the largest
Lyapunov exponent. The importance of insuring ergodicity in molecular
dynamics simulations is illustrated by comparing the results of ergodic
simulations with earlier molecular dynamics simulations.
\end{abstract}
\pacs{{\bf PACS} numbers: }

\section{Introduction}

The simulation of systems having complex potential energy surfaces (PES) 
is often difficult owing to
the problem of quasi-ergodicity.  Quasi-ergodicity can arise in systems
having
several energy minima separated by high energy barriers. When such
situations occur, as for example in proteins, glasses or clusters, 
the system can become
trapped in local basins of the energy landscape, and the ergodic hypothesis
fails on the time scale of the simulation. In the canonical ensemble, the high
energy regions of the PES are exponentially suppressed and 
barrier crossings become rare
events. Calculations of equilibrium properties when phase space is thus
partitioned require methods that overcome quasi-ergodicity by enhanced
barrier crossing. Many techniques have been proposed to address this problem,
including the use of generalized ensembles such as
multicanonical\cite{berg} or Tsallisian,\cite{straub,hansmann}
simulated tempering,\cite{marinari} configurational\cite{frenkel} or
force bias\cite{pangali} Monte Carlo, or various versions of the
jump-walking\cite{frantz90,frantz92,matro,curotto,xuberne}
algorithm. Most of these
techniques have been introduced for Monte Carlo (MC) simulations rather
than
molecular dynamics (MD) simulations. These techniques have been applied to a
variety of sampling and optimization problems, and phase changes in
clusters have often been considered as a benchmark to test these
methods.\cite{straub,curotto,xuberne}

The double-funnel energy landscape of the 38-atom Lennard-Jones (LJ) cluster
has been investigated in detail by Doye, Miller and
Wales,\cite{doye98a,doye98b,doye99,miller99}
who recently also estimated the inter-funnel rate constants using master
equation dynamics.\cite{doye98b} This
landscape is challenging for simulation because of the high free-energy barrier
separating the two funnels.\cite{doye99}
In the preceding paper (hereafter referred to as I),\cite{paperI} 
we have shown how the
parallel tempering algorithm can be used to deal with this particularly
complex system for Monte Carlo simulations in the canonical ensemble.
Achieving ergodicity in microcanonical simulations is much harder than in the
canonical ensemble, because the system is unable to cross any energy
barrier higher than the total energy available.  The 38-atom
Lennard-Jones cluster is fundamentally non-ergodic in a range of energies. 
This non-ergodicity
may not be a serious problem when considering one particular cluster on a short
time scale. However, in a statistical sample of such systems it is
important to have ergodic results.

To allow MD simulations to cross the high energy barriers, one may think of
heating the system (by increasing its kinetic energy), followed by a
cooling back to its initial thermodynamic state. Although this process is
straightforward, the dynamics becomes biased and non physical during the
heating and cooling processes. Moreover, it is difficult to control
accurately the heating and cooling rates that should be chosen for any system.
This latter aspect is particularly critical for the 38-atom Lennard-Jones
cluster where the narrow and deepest funnel is hard to reach even at
high temperatures.

Because of the inherent difficulties of molecular dynamics,
MC approaches can be especially useful for dealing with the problem of
crossing high energy barriers.
Monte Carlo methods have been
developed in previous work\cite{curotto,miller97} in the microcanonical
ensemble.  In these approaches the microcanonical sampling is at fixed
energy without any additional constraints.  Such methods can be
contrasted with isoenergetic molecular dynamics where the total, linear
and angular momenta are also conserved.  These additional
constraints must be considered even at zero angular 
momentum.\cite{dumont,smith,calvo98a}
To differentiate microcanonical simulations, where only the energy is
fixed, from molecular dynamics simulations, where additional
constraints are imposed,
we identify the former to be simulations in the microcanonical ensemble
and identify the latter simulations 
to be in the molecular dynamics ensemble.
The differences in the two ensembles can be particularly important
when the angular momentum is large
enough to induce structural (centrifugal) distortions.\cite{calvo98a}
Because dynamical properties are calculated using molecular dynamics
methods, in this work we find that a combination of Monte Carlo and 
molecular dynamics methods are most convenient for developing ergodic
approaches to dynamics.

In this paper, we adapt the parallel tempering method
to both the microcanonical and molecular dynamics ensembles. The
application of parallel tempering in the molecular dynamics ensemble
requires the incorporation of the
conservation of the total linear and angular momenta into the
probabilities.
In order to extract ergodic dynamical properties,
we combine Monte Carlo methods with molecular dynamics to develop
a hybrid ergodic MC/MD algorithm. The efficiency of the simulation tools
developed in this work is demonstrated by applications 
to the 38-atom Lennard-Jones cluster,
which exhibits a solid-solid transition prior to
melting.\cite{doye98b,paperI}
This transition to
an equilibrium phase between truncated octahedral and icosahedral geometries
makes this cluster an ideal candidate for investigating how the ergodic
hypothesis can influence the dynamical
behavior of a complex system.

The contents of the remainder of this paper are as follows.
In the next section, we recall the basic principle of Monte Carlo sampling in
the microcanonical ensemble, and we present the simple modifications
needed to include parallel tempering. We test 
microcanonical parallel tempering methods on the 38-atom
Lennard-Jones cluster, and compare the microcanonical results with those found
in I using the canonical ensemble. We focus on equilibrium properties, 
including the caloric curve and the
isomers distributions. In section \ref{sec:emd} 
we review the characteristics of
the molecular dynamics ensemble at fixed total linear and angular momenta and
fixed total energy. We extend the parallel tempering Monte Carlo method to the
MD ensemble, and we combine microcanonical
parallel tempering with molecular dynamics to produce an ergodic
MD method. We also apply these methods to several dynamical properties of
LJ$_{38}$; in particular the diffusion constant and the 
largest Lyapunov exponent. We summarize our findings and discuss our
results in
section \ref{sec:ccl}.

\section{Parallel tempering Monte Carlo in the microcanonical ensemble}
\label{sec:ptmc}

The fundamental quantity in the microcanonical ensemble is the density of
states $\Omega$. For a system having $N$ identical particles, volume $V$ and
total energy $E$, $\Omega$ is defined by
\begin{equation}
\Omega(N,V,E) = \frac{1}{N! h^{3N}} \int \delta [H({\bf r},{\bf p}) -E]
d^{3N}r \ d^{3N} p\label{eq:dos}
\end{equation}
where $h$ is Planck's constant and where
$H({\bf r},{\bf p})$ denotes the classical Hamiltonian function of
the coordinates ${\bf r}$ and momenta ${\bf p}$ of the $N$ particles.
Knowing the microcanonical density of states $\Omega$, one can calculate the
canonical partition function $Q(N,V,T)$ by a Laplace 
transformation.\cite{curotto} 
The kinetic part of the Hamiltonian $H$ is
quadratic in the momenta, and Eq. (\ref{eq:dos}) can be partly integrated
to give\cite{curotto,pearson}
\begin{equation}
\Omega (N,V,E) = \left( \frac{2\pi m}{h^2}\right)^{3N/2} \frac{1}{N! \Gamma
(3N/2)} \int \Theta[E-U({\bf r})] [E-U({\bf r})]^{3N/2-1} d^{3N}r.
\label{eq:dosi}
\end{equation}
In Eq.(\ref{eq:dosi}), $\Gamma$ is the Gamma function, $m$ is the mass
of each particle,
$U({\bf r})=H-K$ is the potential energy and $\Theta$ is the
Heaviside step function: $\Theta(x)=1$ if $x\geq 0$, 0 otherwise.
Microcanonical averages of a coordinate-dependent variable $A({\bf r})$ can
be expressed
\begin{equation}
\langle A \rangle (N,V,E) = \frac{\displaystyle \int \Theta[E-U({\bf r})]
[E-U({\bf r})]^{3N/2-1} A({\bf r}) d^{3N}r }{\displaystyle \int
\Theta[E-U({\bf r})] [E-U({\bf r})]^{3N/2-1} d^{3N}r}.
\label{eq:microav}
\end{equation}
The microcanonical
entropy $S$ can be defined by $S(N,V,E) = k_B \ln \Omega(N,V,E)$ with $k_B$
the Boltzmann constant. The thermodynamic temperature $T(N,V,E)$ is
given by the thermodynamic relation
$(\partial S/\partial E)_{N,V} = 1/T$, and can be obtained from a
microcanonical average\cite{pearson}
\begin{equation}
T(N,V,E) = \frac{2}{3N-2} \frac{1}{\langle K^{-1} \rangle}.
\label{eq:tnve}
\end{equation}
This expression is slightly different from the kinetic temperature 
$2\langle K
\rangle/3N$, which is a consequence of our choice in the definition of the
entropy. As discussed by Pearson and co-workers,\cite{pearson}
it is also possible to
define the entropy using the phase space volume 
\begin{equation}
\Phi (N,V,E) = \int_0^E \Omega
(N,V,E') dE'.
\end{equation}
Definitions of the temperature based on $\Omega$ differ from the
temperature based on $\Phi$ to order $1/N$, and the two definitions
agree only in the thermodynamic limit.

Monte Carlo simulations can be used to explore the microcanonical ensemble by
performing a random walk in configuration space. In the standard Metropolis
scheme, a trial move from configuration ${\bf r}_o$ to configuration
${\bf r}_n$ is accepted with the probability\cite{metropolis}
\begin{equation}
{\mathrm acc}({\bf r}_o \to {\bf r}_n ) = \min \left( 1,
\frac{\rho_E({\bf r}_n) T({\bf r}_n \to {\bf r}_o )}
{\rho_E({\bf r}_o) T({\bf r}_o \to {\bf r}_n )}\right ),
\label{eq:metropolis}
\end{equation}
where $T({\bf r}_o \to {\bf r}_n)$ is a trial probability. 
The acceptance probability expressed in Eq.(\ref{eq:metropolis}) insures
detailed balance so that the random walk visits points in configuration
space proportional to the equilibrium
distribution $\rho_E({\bf r})$ defined by
\begin{equation}
\rho_E ({\bf r})
= \zeta^{-1}\Theta[E-U({\bf r})] [E-U({\bf r})]^{3N/2-1}
\end{equation}
where $\zeta$ is the normalization.
In practice, $T({\bf r}_o \to
{\bf r}_n)$ is a uniform distribution of points of width $\Delta$ centered
about ${\bf r}_o$, and $\Delta$ is adjusted as a function of the energy so
that not too many trial moves are either accepted or rejected.

Implementation of microcanonical Monte Carlo is as easy as its
canonical version. Because Monte Carlo methods are based on random walks in
configuration space, in principle the system can
cross energy barriers higher than the available energy. However, in
difficult cases like LJ$_{38}$, large atomic displacements 
having poor acceptance ratios are
needed to reach ergodicity.

Parallel tempering\cite{geyer,tesi,deem,depablo}
has proved to be an important approach to insure ergodicity in canonical
Monte Carlo simulations, and parallel tempering
can be easily adapted to the microcanonical ensemble by
replacing the Boltzmann factors in the acceptance probability by the
microcanonical weight $\rho_E({\bf r})$. In the parallel tempering scheme,
several microcanonical MC simulations are performed simultaneously at
different total energies $\{ E_i\}$. With some predetermined
probability, two simulations
at energies $E_i$ and $E_j$ attempt to exchange their current configurations,
respectively ${\bf r}_i$ and ${\bf r}_j$, and this exchange is accepted with
probability
\[
\min \left( 1,
\frac{\rho_{E_i} ({\bf r}_j ) \rho_{E_j} ({\bf r}_i)}
{\rho_{E_i} ({\bf r}_i ) \rho_{E_j} ({\bf r}_j)} \right) .
\]
The acceptance ratio is analogous to the canonical expression given in
I.  In microcanonical simulations
the potential energies
must be smaller than $\min(E_i,E_j)$; otherwise the move is rejected.
Parallel tempering microcanonical MC works in the same way as in
standard canonical MC.  As with canonical parallel tempering
MC, the gaps between adjacent total energies must be chosen to be small
enough so that exchanges between the corresponding trajectories are accepted
with a reasonable probability.

By using a histogram reweighting technique,\cite{labastie}
it is possible to extract from the
MC simulations the density of states $\Omega$, and then all the thermodynamic
quantities in both the microcanonical and the canonical ensembles. The
procedure is similar to that described in Ref. \CITE{calvo95}, 
and relies on the
calculation of the distribution $P(U,E)$ of potential energy $U$ at the total
energy $E$. $P$ is fitted to the microcanonical form $P(U,E) = \Omega_C(U)
(E-U)^{3N/2-1}/\Omega(E)$, where  $\Omega_C$ stands for the configurational
density of states, and $\Omega(E)$ is extracted by convolution of $\Omega_C
(U)$ and $(E-U)^{3N/2-1}$.

We have tested the parallel tempering Monte Carlo algorithm in the
microcanonical ensemble on the 38-atom Lennard-Jones cluster previously
investigated. Forty 
different total energies ranging from $-172.4737\varepsilon$
to $-124\varepsilon$ have been used, and 
the same simulation conditions have been chosen
as in I.
In addition to a constraining sphere of radius
$2.25\sigma$ to prevent evaporation, exchanges have been attempted every
10 passes,
with the same method for choosing exchanging trajectories
as described in the previous article.
The simulations are begun with random configurations of the
cluster geometry, and consist of $1.3\times 10^{10}$ points accumulated
following equilibration moves consisting of $95\times 10^6$ 
Metropolis points(no exchanges) followed by
$190\times 10^6$ points using parallel tempering.  The microcanonical
heat capacity calculated in this fashion and shown in Fig. \ref{fig:microcv},
is qualitatively the same as
the canonical heat capacity [see I].  The melting peak in the microcanonical
heat capacity occurs at the same calcuated temperature as the
temperature of the melting peak in the
canonical heat capacity, and there
are slope change regions at temperatures that correspond to
equilibrium between the icosahedral basin and the truncated octahedral
global minimum structure.
The present simulations are also used to obtain structural
insight about the cluster as a function of total energy. 
We have calculated the
order parameter $Q_4$ as defined in I as a function of temperature, and
compared the classification into the three categories of isomers
(truncated
octahedral, icosahedral or
liquid-like) using the energy criterion also outlined in I.

In Fig. \ref{fig:caloric} we show the caloric curve $T(E)$
determined from our parallel tempering microcanonical MC simulations. We
also present the canonical curve for comparison. The melting transition
near $T\sim 0.166\varepsilon/k_B$ is reflected in the change in slope of the
temperature as a function of the energy.
The microcanonical curve does not display a van der Waals loop,
and remains very close to the canonical curve. The average value of the order
parameter $\langle Q_4 \rangle$ is displayed in the lower panel of Fig.
\ref{fig:caloric} as a function of the total energy. 
As has been discussed in I for the canonical simulation,
the order parameter begins to drop at energies where there is the onset
of isomerization transitions to the icosahedral basin
(near $E=-160\varepsilon$), and the order parameter reaches its
lowest value at the melting transition. The isomer distributions have been 
evaluated either
using the parameter $Q_4$ 
or using the energy criterion (see the discussion in paper I). 
The results have been plotted in Fig. \ref{fig:str}
as a function of the total energy. The behavior of the isomer
distributions as a function of energy is similar to the
canonical distributions as a function of temperature, and the cluster 
exhibits
equilibrium between truncated octahedral and icosahedral geometries in the
energy range $-160\varepsilon \lesssim E \lesssim -150\varepsilon$, prior to
the solid-like to liquid-like phase change. As in the canonical case, the
icosahedral distribution is a symmetric function of the energy 
when the energy criterion is used rather than the definition based on
$Q_4$.
This difference reflects the differences between the two definitions
of icosahedral and liquid basins.  The oscillatory structure observed at
the peak of $P_{Q_4}$ for the icosahedral distribution in the upper
panel of Fig. \ref{fig:str} is smaller than the calculated errors (two
standard deviations of the mean are shown).  Whether the observed
structure would persist for a longer simulation is not known to us. 
Because the definition that assigns configurations to the icosahedral
basin is arbitrary, we have chosen not to investigate this structure
further.

It is useful to contrast the current 
results with previous constant energy studies of
LJ$_{38}$.  Previous simulations have used molecular dynamics methods where
no attempt has been made to insure ergodicity.  To contrast these past
studies with the molecular dynamics technique discussed in the next
section of this paper, we define {\em standard molecular dynamics\/} to
represent the usual molecular dynamics method where no special procedure
is introduced to insure ergodicity.
Simulations of LJ$_{38}$ using standard molecular dynamics invariably lead to a
caloric curve with a clear van der Waals loop and a
melting temperature higher than that inferred from Fig. 
\ref{fig:caloric}.\cite{calvo98b} From the results of Ref. \CITE{calvo98b}, 
the cluster is
trapped in the octahedral basin, and the system does reflect the true
dynamical coexistence state
between the truncated octahedron and the icosahedral
basin. This is the common situation encountered in MD simulations of the
LJ$_{38}$ system; the cluster chooses either to remain trapped in the
octahedral
basin, or to escape and coexist between the icosahedral solid-like
and liquid-like forms. Because the system is unable to return from the
octahedral basin,
the microcanonical temperature decreases.
In the usual case, van der Waals loops arise when there are large
energy gaps between the lowest-energy isomers.\cite{wales93}
In the specific case of LJ$_{38}$, it appears that the presence of extra
(icosahedral) isomers only slightly
higher in energy than the octahedral structure eliminates this
loop in the ergodic microcanonical caloric curve.

In order to extract dynamical quantities, the Monte
Carlo method we have presented must be modified to sample the MD
ensemble.  The modification is the subject of the next section.

\section{Ergodic molecular dynamics}
\label{sec:emd}

The molecular dynamics ensemble differs from the microcanonical ensemble in
that two quantities are conserved in addition to the total energy $E$,
volume $V$ and number of particles $N$. These two quantities are the
total linear momentum ${\bf P}$ and total angular momentum ${\bf L}$. If
their values are prescribed, the density of states remains the fundamental
quantity of interest, and is now defined by
\begin{eqnarray}
&&\Omega(N,V,E,{\bf P},{\bf L}) = \nonumber \\
&&\frac{1}{N! h^{3N}} \int \delta [H({\bf r},{\bf p})-E] \delta \left( {\bf P}
- \sum_{i=1}^N {\bf p}_i \right) \delta \left( {\bf L} 
- \sum_{i=1}^{N} {\bf r}_i\times {\bf p}_i
\right) d^{3N}r d^{3N}p.
\label{eq:dostotal}
\end{eqnarray}
As is the case in the microcanonical ensemble [see Eq.(\ref{eq:dosi})],
for atomic systems the momentum integrations in Eq.(\ref{eq:dostotal})
can be evaluated explicitly.\cite{dumont,smith,calvo98a}
Because the thermodynamic properties are not 
affected by the translational motion
of the center of mass, we can assume that ${\bf P} =0$. We then
obtain\cite{calvo98a}
\begin{eqnarray}
&&\Omega(N,V,E,{\bf P}=0,{\bf L}) = \nonumber \\
&&\left( \frac{2\pi m}{h^2}\right)^{3N/2-3} \frac{1}{N! \Gamma(3N/2-3)} \int
\Theta[E-U_{\bf L}({\bf r})] [E-U_{\bf L}({\bf r})]^{3N/2-4} \frac{d^{3N}
r}{\sqrt{{\mathrm det}\,{\bf I}}},
\label{eq:dosl}
\end{eqnarray}
where ${\bf I}$ is the inertia matrix and $U_{\bf L}({\bf r}) = U({\bf r}) +
{\bf L}^\dagger{\bf I}^{-1}{\bf L}/2$ is the effective rovibrational energy.
This effective potential energy includes the kinetic energy contribution of the
rotating cluster considered as a rigid body.\cite{jellinek89a,jellinek89b}
The landscape of rotating clusters
has been investigated by Miller and Wales in order to study cluster
evaporation.\cite{miller96} Averages in the MD ensemble are now expressed as
\begin{equation}
\langle A \rangle = \frac{\displaystyle \int \Theta[E-U_{\bf L}({\bf r})]
[E-U_{\bf L}({\bf r})]^{3N/2-4} A({\bf r})
\frac{d^{3N} r}{\sqrt{{\mathrm det}\,{\bf I}}}}
{\displaystyle \int \Theta[E-U_{\bf L}({\bf r})]
[E-U_{\bf L}({\bf r})]^{3N/2-4}
\frac{d^{3N} r}{\sqrt{{\mathrm det}\,{\bf I}}}}.
\label{eq:mdav}
\end{equation}
As in the microcanonical ensemble, we define the entropy in the
molecular dynamics ensemble by
$S=k_B \ln \Omega$. The differences
between the microcanonical and molecular dynamics ensembles are the exponent
$3N/2$ which is reduced by 3 owing to the 
geometrical constraints, the potential
energy which now includes the contribution of the centrifugal energy, and the
weight $1/\sqrt{{\mathrm det}\,{\bf I}}$ which 
is a consequence of the conservation
of the angular momentum.
Monte Carlo simulations can sample the MD ensemble by performing
a random walk in configuration space. The acceptance probability from
configuration ${\bf r}_o$ to configuration ${\bf r}_n$ is
\begin{equation}
{\mathrm acc}({\bf r}_o \to {\bf r}_n ) = \min \left( 1,
\frac{\rho_{E,{\bf L}} ({\bf r}_n) T({\bf r}_n \to {\bf r}_o)}
{\rho_{E,{\bf L}} ({\bf r}_o) T({\bf r}_o \to {\bf r}_n)} \right)
\label{eq:accmde}
\end{equation}
in the Metropolis scheme. The microcanonical weight $\rho_E({\bf r})$ is now
replaced by the MD weight $\rho_{E,{\bf L}}$ given by
\begin{equation}
\rho_{E,{\bf L}}({\bf r}) = \zeta^{-1}\frac{1}{\sqrt{{\mathrm det}\,{\bf I}}}
\Theta[E-U_{\bf L}({\bf r})][E-U_{\bf L}({\bf r})]^{3N/2-4},
\label{eq:nverho}
\end{equation}
where $\zeta$ is a normalization.
The expression for the acceptance probability is similar to Eq.
(\ref{eq:metropolis}), and a practical implementation of Monte Carlo in the
MD ensemble is made in the same way as in the true microcanonical ensemble,
given the vector ${\bf L}$. Parallel tempering can be also easily combined
with the MC simulations. The acceptance probability of exchanging the
configurations ${\bf r}_i$ and ${\bf r}_j$ initially at the total energies
$E_i$ and $E_j$ respectively is then
\[
\min \left( 1,
\left( \frac{[E_i-U_{\bf L}({\bf r}_j)][E_j-U_{\bf L}({\bf r}_i)]}
{[E_i-U_{\bf L}({\bf r}_i)][E_j-U_{\bf L}({\bf r}_j)]} \right)^{3N/2-4}
\right)
\]
provided that all quantities inside brackets are positive (otherwise the move
is rejected). It is remarkable that the geometrical weights have canceled
in this expression.

The Monte Carlo method we have just described allows sampling of
configuration space rigorously equivalent to the sampling we would obtain using
molecular dynamics, but with the additional possibility of crossing the energy
barriers higher than the available energy. The method 
can be used in its present form
to extract equilibrium properties only dependent on the energy or
geometry, as has been illustrated in the 
previous section. To compute dynamical
quantities, the method can also provide a 
database of configurations representative of
a given total energy. Instead of performing a few very long MD
simulations that are in principle unable to reach other parts of the energy
surface separated by barriers higher than the available energy, we choose to
perform a statistical number of short simulations starting from configurations
obtained by parallel tempering Monte Carlo in the MD ensemble with same total
energy and angular momentum. By construction, if the MC method is correctly
ergodic, then the hybrid MD method we have suggested can be expected to yield
ergodic dynamical observables.

We now illustrate this ergodic molecular dynamics method on the LJ$_{38}$
problem. Two essentially dynamical parameters have been calculated. The first
is the self diffusion constant $D$, obtained from the derivative of the
average mean square atomic displacement
\begin{equation}
D = \frac{1}{6} \frac{d}{dt} \langle [{\bf r}(t) - {\bf r}(0)]^2\rangle,
\label{eq:dif}
\end{equation}
where the average is taken over all particles of the system and over all short
MD simulations.
The other parameter is the largest Lyapunov exponent $\lambda$, that measures
the exponential rate of divergence of the distance between two initially close
trajectories in the phase space. If we write the equation describing the
Hamiltonian dynamics in condensed form as $\dot{\psi}(t)=F(\psi)$ where $F$ is a
nonlinear function and $\psi=\{ {\bf r}, {\bf p} \}$ the phase space point,
then a small perturbation $\delta \psi$ evolves according to the simple
equation $d\delta \psi/dt = 
(\partial F/\partial \psi)\delta \psi$. The
largest Lyapunov exponent $\lambda$ is obtained from the time evolution of the
vector $\delta \psi$:
\begin{equation}
\lambda = \lim_{t\to \infty} \lim_{\delta \psi (0) \to 0} \frac{1}{t} \ln
\frac{\| \delta \psi (t) \|}{\| \delta \psi (0) \|}.
\label{eq:mle}
\end{equation}
In Eq.(\ref{eq:mle}), $\| \cdot \|$ is a 
metric on the phase space. In principle,
any metric can be used, and we choose the Euclidian metric including both the
momenta and the coordinates. The numerical
procedure\cite{benettin}
involves a periodic renormalization of the vector $\delta \psi$ to
prevent its exponential divergence. The successive lengths are accumulated and
contribute to the average value of $\lambda$.

In I, the clusters have been defined using a hard sphere constraining
potential.  Because 
the angular momentum is not conserved after reflection from such hard wall
boundaries, in the molecular dynamics simulations
we have chosen a soft
repulsive spherical wall $U_c$ defined with respect to the center
of mass of the cluster for each particle by
\begin{equation}
U_c({\bf r}) = \cases{0, & $r < R_c$ \cr \kappa(r-R_c)^4/4, &
$r\geq R_c$.}
\label{eq:vrep}
\end{equation}
In this equation, the atomic distances $r$ are measured with respect to the
cluster center of mass. The simulations have been 
performed setting the angular momentum to zero for simplicity. 
We stress that even in this case (with ${\bf L}=0$), the weight $1/
\sqrt{{\mathrm det}\,{\bf I}}$ must be included in the Monte Carlo
probabilities so that we effectively sample the MD ensemble. 
The actual thermodynamic behavior in the MD ensemble at zero angular
momentum is nevertheless nearly identical to the microcanonical behavior.

The application to the LJ$_{38}$ cluster has been made by performing $10^{10}$
MC steps following $10^7$ equilibration steps in a parallel tempering
simulation in the MD ensemble. The same 40 total energies have been 
chosen as in the
previous section, and $10^5$ configurations have been 
stored every $10^5$ steps for
each simulation. Short molecular dynamics runs of $10^4$ time steps
following $10^3$ equilibration steps have been performed for each of these
configurations, with the same total energy as the corresponding MC trajectory
of origin, and with zero total linear and angular momenta as well.
The parameters used
for the constraining wall are respectively $R_c=2.25\sigma$ and 
$\kappa=100\varepsilon$,
for both the MC and MD runs. A simple Verlet
algorithm has been 
used to propagate the MD trajectory with the time step $\delta
t=0.01$ reduced LJ units. The propagation of the tangent trajectory to
calculate the Lyapunov exponent has been determined 
with a fourth order Runge-Kutta
scheme. The final values of $D$ and $\lambda$ are an average over the $10^5$
MD simulations. The variations of $D$ and $\lambda$ with total energy are
depicted in Fig. \ref{fig:dl}. In both cases, two curves have been plotted,
calculated either from standard molecular dynamics (with $10^8$ time steps
following $10^7$ equilibration steps, and starting initially from the
lowest-energy structure), or from our hybrid ergodic
molecular dynamics method. For both quantities, the two MD schemes clearly
yield distinct values in the energy range where equilibrium between truncated
octahedral and icosahedral geometries occurs. The thermodynamic temperature,
not plotted here, has the same variations as the caloric curve of Fig.
\ref{fig:caloric} when calculated with ergodic MD. Standard molecular
dynamics predicts a
van der Waals loop centered at $T\sim 0.18 \varepsilon/k_B$. For
standard MD,
the cluster is trapped in the icosahedral basin and is, in practice,
unable to reach the octahedral basin. Only the
equilibrium between the icosahedral basin and liquid-like structures
occurs.
As can be seen from the upper panel of Fig. \ref{fig:dl}, this change in
curvature of the temperature is also present for the diffusion constant, which
exhibits strong variations at the energy where the 
octahedral structure vanishes
when standard MD is used. In contrast, the variations in ergodic
MD are smooth.

The melting temperature implied by the largest
Lyapunov exponent is also higher in standard MD than in ergodic MD,
even though the variations of the Lyapunov exponent
are continuous in both
MD schemes.\cite{calvo98b}
Indeed, using ergodic molecular dynamics we observe a shift of the
curve obtained by standard MD toward the lower energies. As shown by Hinde,
Berry, and Wales,\cite{hinde92,hinde93,wales91}
the Lyapunov exponent and the Kolmogorov entropy are
quantities essentially dependent on the local properties of the energy
landscapes. One contribution comes from the negative curvature of the
landscape, and another contribution is the fluctuation of positive
curvature.\cite{mehra}
Both contributions are affected by the cluster being trapped either inside
the truncated octahedral basin or inside the icosahedral basin. In this latter
case in particular, the different isomers belonging to the icosahedral
basin are connected through regions of negative curvature, while only one
isomer defines the octahedral funnel.

Because ergodic molecular dynamics allows the cluster to be found in both
basins prior to melting, the dynamical behavior is likely to be very
different (and more chaotic) with respect to the dynamical behavior of the
cluster confined to the octahedral funnel. This
difference is precisely what we observe on the
lower panel of Fig. \ref{fig:dl}.

\section{Conclusion}
\label{sec:ccl}

In this paper, we have explored the parallel tempering method in
simulations in the microcanonical ensemble. The implementation of the parallel
tempering algorithm in this ensemble is straightforward,
the Boltzmann factor $\exp (-\beta U)$ being replaced by the microcanonical
weight $(E-U)^{3N/2-1}$. Application to the LJ$_{38}$ cluster has shown
the thermodynamic behavior in the microcanonical ensemble to be similar to the
behavior in the canonical ensemble. The solid-liquid phase change is preceded
by a solid-solid phase change where the cluster is in equilibrium between
truncated octahedral and icosahedral geometries. This phase equilibrium is well
reproduced in the simulations owing to the power of parallel tempering.
The calculated microcanonical caloric curve, which does not display a van
der Waals loop, is consistent with the single peaked heat capacity observed
in I.\cite{paperI}

We have extended the parallel tempering microcanonical Monte Carlo algorithm
to sample the molecular dynamics ensemble at constant total energy, linear
momentum and angular momentum. Combined with standard molecular dynamics, this
method circumvents the lack of connectivity between regions of the
potential energy surface. The method can ensure ergodicity in microcanonical
simulations, which is much more difficult to achieve than in the canonical
ensemble. Ironically, this ergodic MD method can be viewed as the counterpart
of the techniques developed by Chekmarev and Krivov to study the dynamics
of systems confined to only one catchment basin in the energy
surface.\cite{chekmarev}

We have performed a statistical number of short molecular dynamics runs
starting from configurations stored periodically in parallel tempering Monte
Carlo simulations. These simulations sample the MD ensemble at the same total
energies, linear and angular momenta as the standard 
molecular dynamics runs. In
fact, the length of the MD runs is mainly dictated by the large number of
starting configurations. One may think of reducing drastically this number, to
allow for the calculation of parameters varying on longer time scales.
Unfortunately, if the energy landscape is not known in advance, then it
is hard to guess how important are the contributions of the basins not
selected as starting configurations. In the case of LJ$_{38}$ having only 3
main regions on the energy surface, one possibility is to compute a dynamical
property as the average value over 3 different simulations starting either from
the truncated octahedral geometry, one icosahedral geometry or a low-lying
liquid geometry, all carried out at the same total energy. However, as we have
seen in Fig. \ref{fig:str}, it is not obvious how 
to choose properly the weights
of each basin in this average because of the difficulty in distinguishing 
between
icosahedral and liquid structures in many cases. For this reason, we believe
that the first parallel tempering
MC step of the hybrid ergodic method is essential in the
vicinity of phase changes to capture many starting configurations that
are used subsequently in standard molecular dynamics. The enhanced sampling
offered by parallel tempering can also act as a statistical representation
of the energy surface at a given total energy, and the long time dynamics
may be further investigated by using master equations after searching the
saddle points.\cite{miller99,ball}

We have calculated two dynamical quantities with the present
hybrid MD/MC method,
the diffusion constant and the largest Lyapunov exponent in the 38-atom
Lennard-Jones cluster. The variations of both quantities with the total
energy are significantly different when evaluated with standard (non-ergodic)
molecular dynamics or with our hybrid ergodic MD method. These results
emphasize the different contributions of the two funnels of the energy
landscape to the average value of the parameters estimated.

The algorithms developed in this investigation allow the calculation of
thermodynamic, structural, or dynamical properties of systems such as
LJ$_{38}$ that can be expressed as phase space or time averages. Parallel
tempering works using a criterion based on the potential energy but not on
the geometry. Consequently permutational isomers
can be introduced in the course of the simulation. Quantities
such as fluctuations of configuration-dependent properties are much more
difficult to extract than actual averages. For instance, the Lindemann index
$\delta$, which measures the root mean square bond length fluctuation, is
often considered to be 
a reliable parameter for detecting melting in atomic and
molecular systems. This quantity cannot
be properly estimated with the ergodic MD scheme, and the same difficulty
persists for other methods based on the use of different trajectories.

Although the idea of combining Monte Carlo sampling with standard 
molecular dynamics
can be applied to other techniques such as jump-walking, we believe that
parallel tempering is the key to the success in the case of LJ$_{38}$. As in
the canonical version, the equilibrium phase between truncated octahedral
and icosahedral structures is correctly reproduced in an energy range preceding
the melting region,
because in this range configurations may be accessed either from higher energy
trajectories containing mainly icosahedral geometries, or from lower energy
trajectories acting as a reservoir for the 
octahedral geometry. As noticed by Falcioni
and Deem,\cite{deem}
the parallel tempering algorithm is especially useful at low
temperatures, or in our case, at low energy. The long relaxation times
inherent in systems like clusters, proteins, critical or glassy
liquids, are a serious difficulty for standard simulation methods. We expect
the present ergodic method to be particularly useful to deal with the dynamics
of such systems.

The method we have presented works at constant
total energy. It is possible to improve ergodicity in constant-temperature MD
either by using canonical parallel tempering as in the work of Sugita
and Okamoto\cite{okamoto}, or
by coupling parallel tempering canonical Monte Carlo to short Nos\'e-Hoover
trajectories. In the Nos\'e-Hoover approach 
such molecular dynamics simulations do conserve a zero
angular momentum, so a rigorous MC sampling should include the geometrical
weight $1/\sqrt{{\mathrm det}\,{\bf I}}$ in the probabilities also in this
case. The present microcanonical scheme can be easily used for rotating bodies,
which makes the method suitable for investigating the strong influence of
centrifugal effects on phase changes in atomic clusters.

\section*{Acknowledgments}

Some of this work has been 
motivated by the attendance of two of us (DLF and FC)
at a recent CECAM meeting on `Overcoming broken ergodicity in simulations of
                    condensed matter systems.' 
We would like to thank CECAM, J.E. Straub
and B. Smit who organized the meeting, and those who attended the
workshop for stimulating discussions.  This work has
been supported in part by the National Science Foundation under grant numbers
CHE-9714970 and CDA-9724347. This research has been supported in part by the
Phillips Laboratory, Air Force Material Command, USAF, through the use of the
MHPCC under cooperative agreement number F29601-93-0001. The views and
conclusions contained in this document are those of the authors and should not
be interpreted as necessarily representing the official policies or
endorsements, either expressed or implied, of Phillips Laboratory or the US
Government.

\begin{figure}
\caption{The heat capacity as a function of energy calculated in the
microcanonical ensemble.  The melting peak occurs at the same
calculated temperature in the microcanonical ensemble as found in the
canonical ensemble, but the height of the microcanonical peak
is significantly higher than the canonical peak [compare with Fig. 1 in
I].  Both the microcanonical and canonical heat capacities display a
region of change in slope at the transition between the truncated
octahedron and the icosahedral basin.  The error bars represent two
standard deviations of the mean.}
\label{fig:microcv}
\end{figure}

\begin{figure}
\caption{Upper panel:
the microcanonical caloric curve for LJ$_{38}$ obtained from parallel
tempering Monte Carlo simulations. The temperature is plotted as a function of
the total energy, both expressed in reduced LJ units. The circles
are the direct results of microcanonical simulations.
The solid line is a fit obtained
by the histogram reweighting technique. Also plotted as a dotted line is the
caloric curve in the canonical ensemble. Lower panel: average value of the
order parameter $Q_4$ as a funciton of the total energy. For both panels, the
error bars are smaller than the size of the symbols.}
\label{fig:caloric}
\end{figure}

\begin{figure}
\caption{Upper panel:
the probability distribution of the order parameter $Q_4$ as a
function of the total energy. Lower panel: the probability distribution of
the energy of the quenched structure as a function of the total energy. For
both quantities, FCC labels the truncated octahedron, IC labels structures
from the icosahedral basin and LIQ labels structures from the liquid region.
In the lower panel, the error bars are smaller than the size of the
symbols.  In the upper panel, the error bars represent two standard
deviations of the mean.}
\label{fig:str}
\end{figure}

\begin{figure}
\caption{Two dynamical parameters calculated for LJ$_{38}$ using either
standard molecular dynamics starting from the lowest-energy structure (empty
symbols) or the hybrid ergodic MD/MC method (full symbols), as a function of
the total energy. The results are expressed in Lennard-Jones time units $t_0$.
Upper panel: diffusion constant $D$; lower panel: largest
Lyapunov exponent $\lambda$.  For both panels, the
error bars are smaller than the size of the symbols.}
\label{fig:dl}
\end{figure}
\newpage
\centerline{\psfig{figure=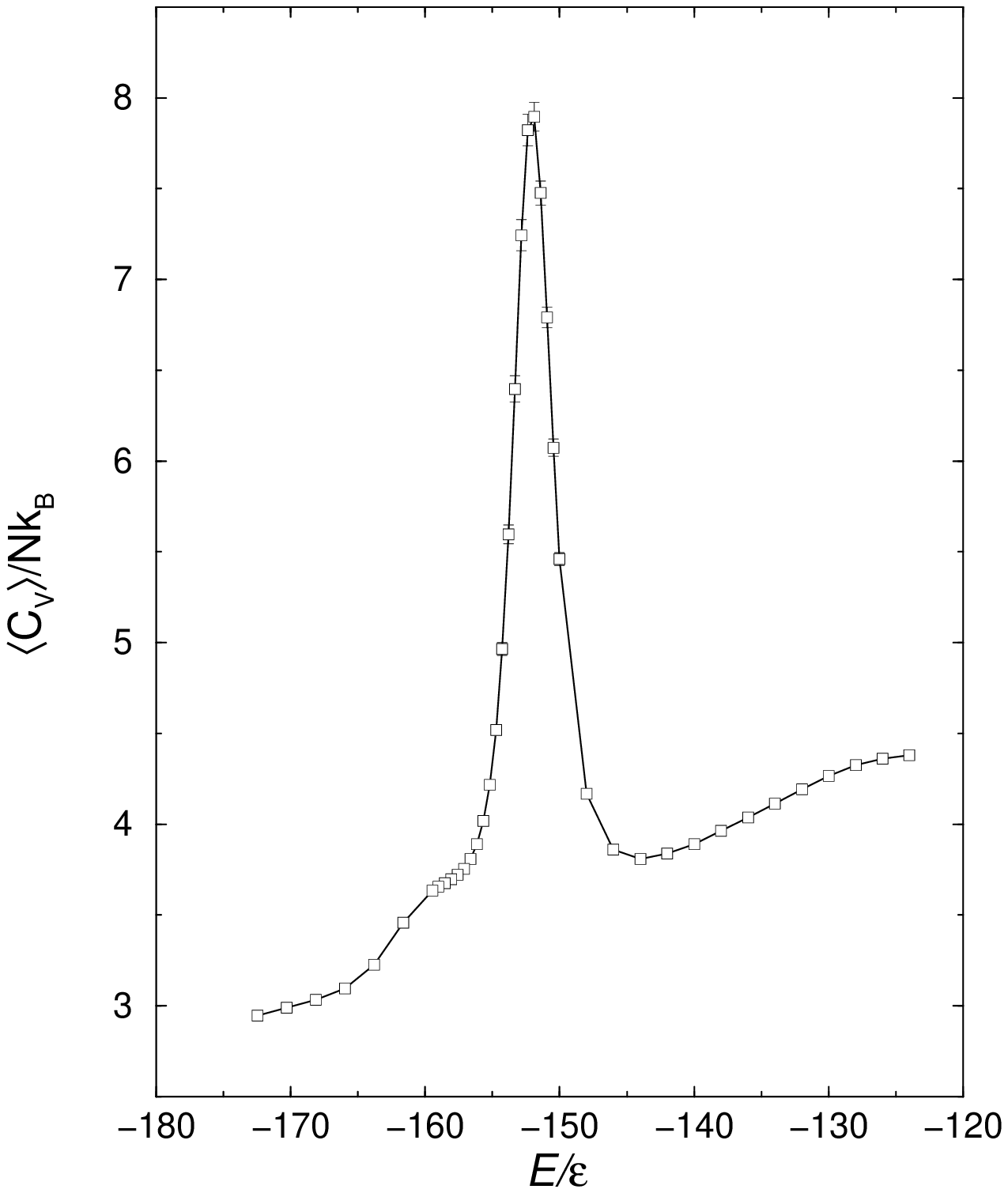}}
\newpage
\centerline{\psfig{figure=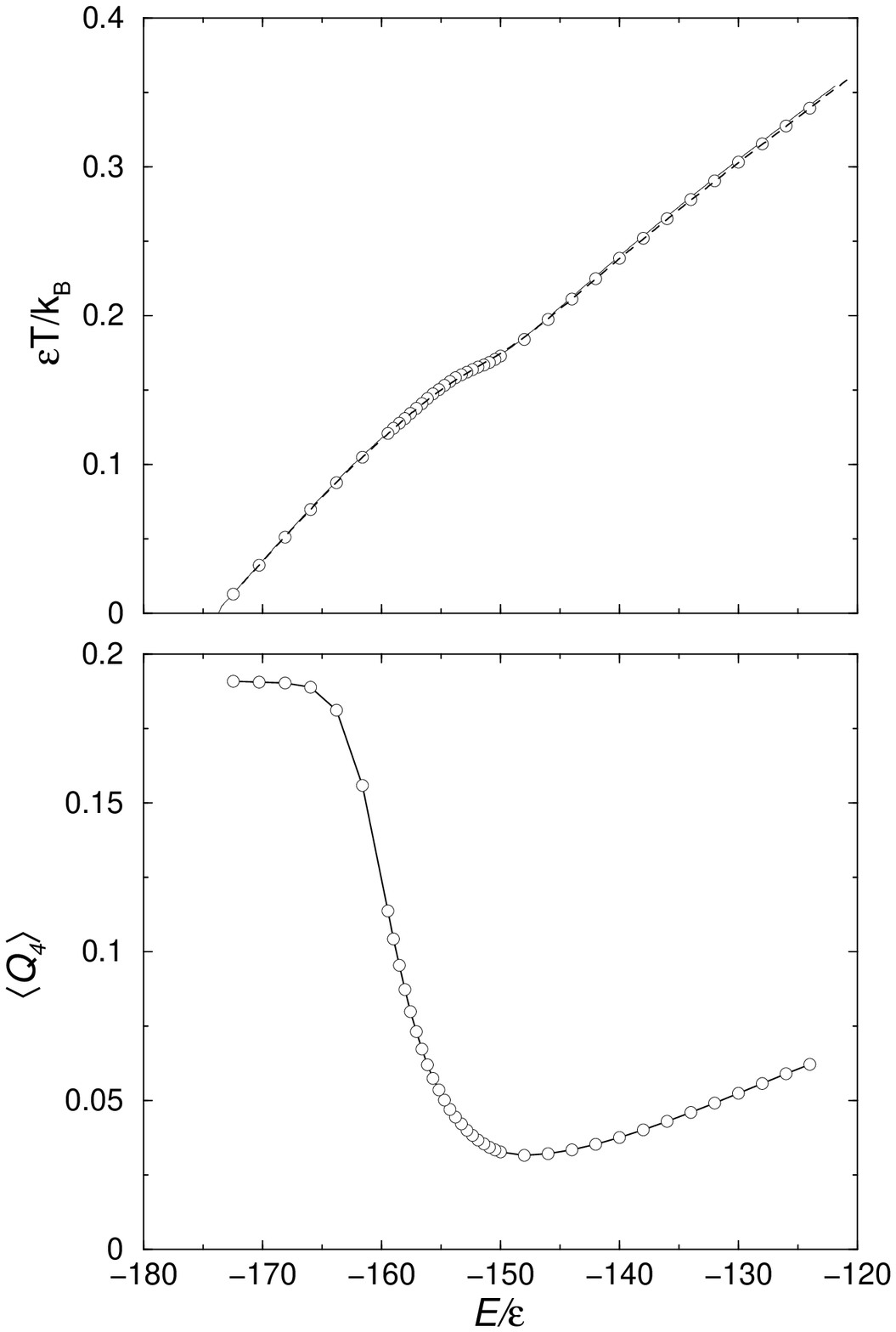}}
\newpage
\centerline{\psfig{figure=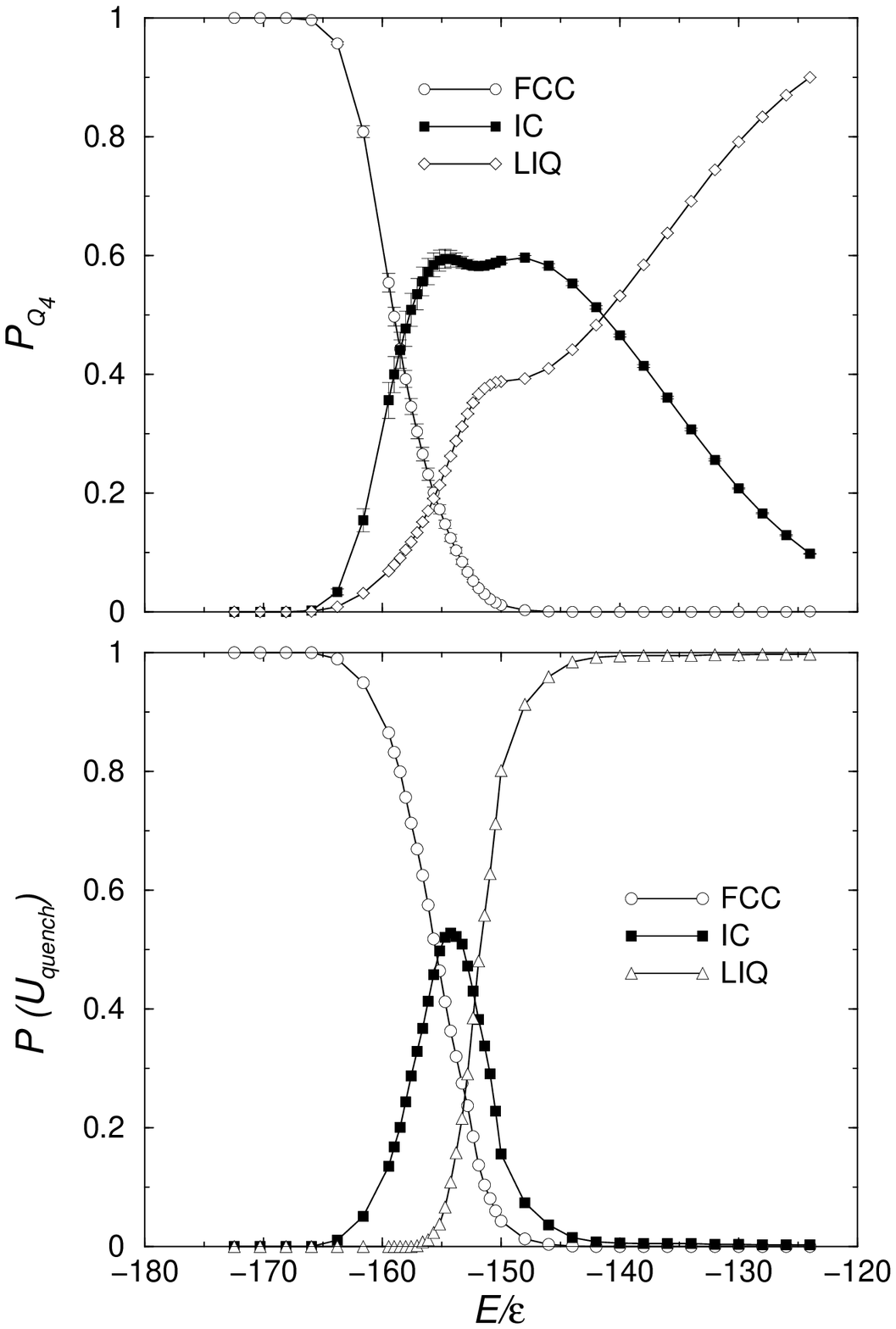}}
\newpage
\centerline{\psfig{figure=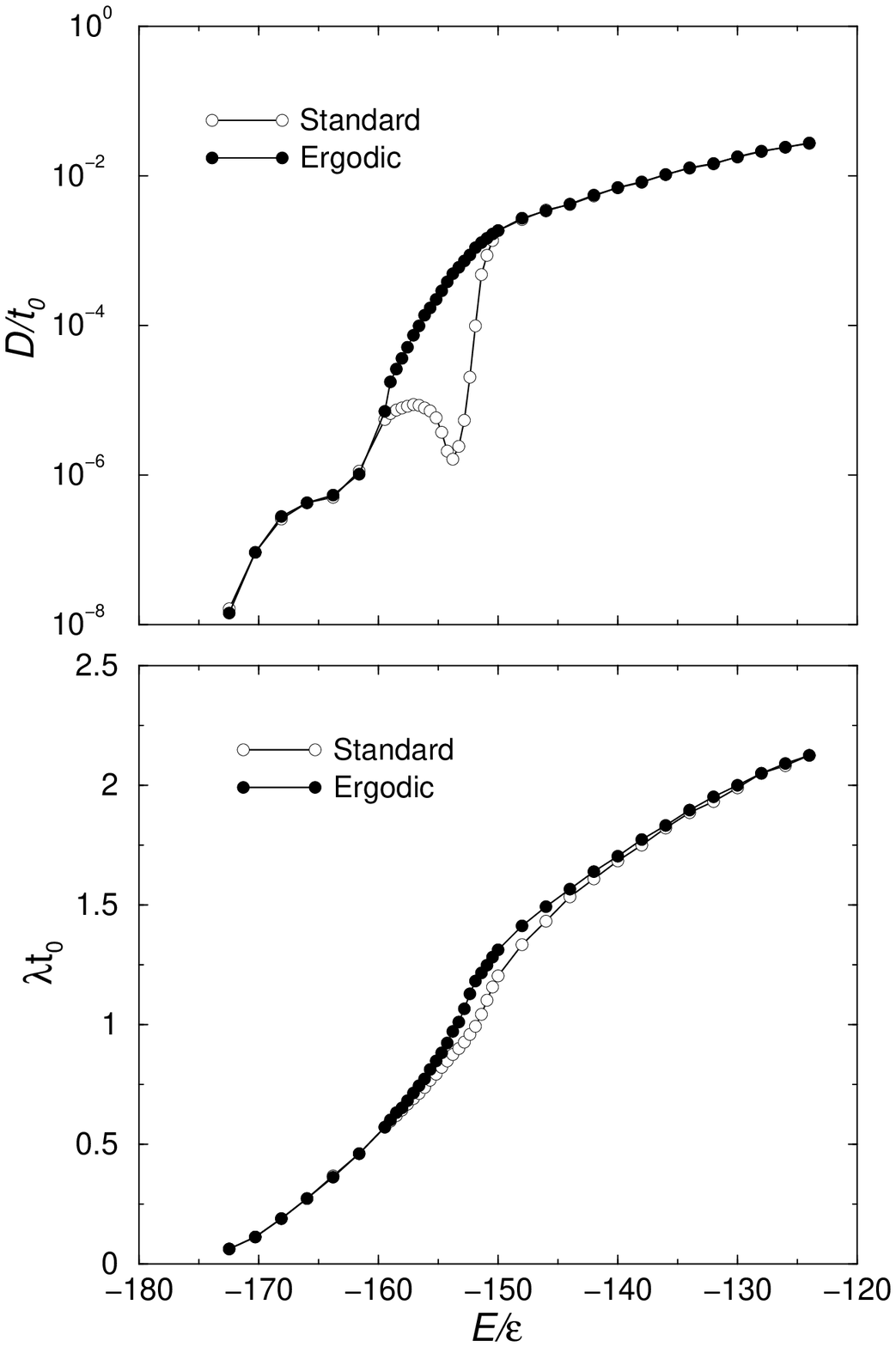}}
\end{document}